# The SPIRIT Telescope Initiative: six years on

*Paul Luckas*
*International Centre for Radio Astronomy Research, The University of Western Australia*
*35 Stirling Hwy, Crawley, Western Australia 6009*
*paul.luckas@uwa.edu.au*

**Abstract**

Now in its sixth year of operation, the SPIRIT initiative remains unique in Australia, as a robust web-enabled robotic telescope initiative funded for education and outreach. With multiple modes of operation catering for a variety of usage scenarios *and* a fully supported education program, SPIRIT provides free access to contemporary astronomical tools for students and educators in Western Australia and beyond. The technical solution itself provides an excellent model for low cost robotic telescope installations, and the education program has evolved over time to include a broad range of student experiences—from engagement activities to authentic science. This paper details the robotic telescope solution, student interface and educational philosophy, summarises achievements and lessons learned and examines the possibilities for future enhancement including spectroscopy.

## 1. Introduction

It is a reality that much of modern observational astronomy is undertaken using highly automated telescopes. Situated often in remote locations, these instruments provide researchers the means to acquire high quality data with great efficiently. Robotic minor planet survey telescopes map the sky adding dozens of new discoveries each week, and satellites command ground based telescope systems to respond rapidly to distant transient phenomena such as gamma ray bursts. Sensitive digital imagers and sophisticated processing software have long replaced the chemical emulsions and analogue analysis of days' past and professional astronomers are now served astronomical data via the internet from the comfort of their campus offices. Yet despite this renaissance, high school text books still often characterise the *astronomer* as someone who *looks* though a telescope—with many that portray astronomy with a picture of Galileo alongside his primitive telescope 400 years ago. Dispelling these notions and providing school students access to the tools used by modern observational astronomers has formed the core of a program of activities within the SPIRIT initiative at The University of Western Australia (UWA) for the past 6 years. Commissioned in 2010 and followed by a second instrument in 2012, SPIRIT comprises two web enabled robotic telescopes located on the roof of the School of Physics at UWA. The initiative is currently hosted by the International Centre for Radio Astronomy Research (ICRAR) as part of its outreach and education program and includes a full life cycle of workshops and student activities.

The provision of robotic telescopes for use in education is not a new. Seminal projects such as the Bradford Robotic Telescope, the Faulke's telescope project, and in more recent years the Las Cumbres global telescope network among many others have done much to increase student participation in contemporary astronomy. Some of these examples are characterised by large observatory-class instruments that, while delivering research quality data potential, incorporate significant operational and maintenance overhead and commensurate operating budgets. In addition, the proprietary nature of their technical solutions (often a consequence of pre-existing observatory infrastructure) presents most as poor models for replication. At the same time, small robotic telescope technology has advanced considerably, evidenced most notably in its use among the amateur community. Advanced amateurs have increasingly made significant contributions to astronomical research in areas such as the discovery and monitoring of transient events—supernovae surveys, minor planet astrometry—as well as photometric measurements of variable stars, exoplanet research and spectroscopy. As a consequence of being part-time astronomers, many have found themselves at the forefront of telescope automation in order to improve output while balancing other 'life commitments'. This is exemplified by the number of vendors of commercial telescope hardware and software who are themselves, amateur astronomers. The small robotic telescope revolution of recent years has impacted not just professional and amateur astronomy, but also education and outreach, particularly in regions where access to established observatory outreach programs is otherwise absent.

Of critical importance has been the realisation that *just* providing access to robotic telescopes does not guarantee that they will be used effectively in science education. Without access to relevant and engaging educational resources, many except for the very enthusiastic of teachers may be reluctant to incorporate the use of such telescopes in their day to day teaching—particularly when the curriculum does not prescribe it. Supporting programs that include teacher workshops and the provision of student activities often typify aspects of a remote telescope initiative that are yet to be fully realised, as is ongoing study and research into the educational effectiveness of the individual programs themselves.

The SPIRIT initiative began with a set of goals that informed a proposal and design in early 2009. Since that time, SPIRIT I & II have proven themselves successful models for modest sized and comparatively low cost web-enabled telescopes exhibiting high levels of reliability. The supporting education program, initially developed through a collaboration between the Western Australian Department of Education and UWA has evolved to include opportunities for students to participate in genuine science.

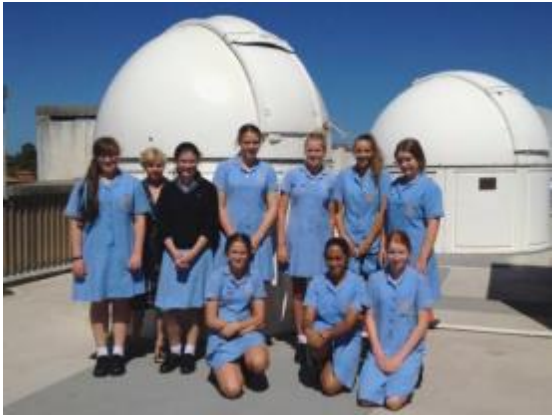

**Figure 1: Students visiting SPIRIT I & II observatories at UWA.**

## 2. Technical Solution

*Vendor information has been omitted from the text in this section for reasons of clarity but can be found in tables at the end of this paper and in the references.*

### 2.1 Hardware

The SPIRIT telescopes are located on the roof of UWA's School of Physics, housed within fibreglass domes manufactured by Sirius Observatories. Dome rotation and shutter control are fully automated utilising the MaxDome interface, and each includes automated and redundant weather monitoring via a Boltwood cloud sensor located adjacent to each observatory. Both optical tube assemblies are mounted on Paramount ME robotic mounts manufactured by Software Bisque. These mounts deliver robust targeting and tracking as well as the capability for remote initialisation via a proprietary homing feature. The Paramount robotic telescope mounts also include advanced periodic error correction, pointing and tracking algorithms through included software. As a consequence, the SPIRIT telescopes do not use auto-guiders—improving the efficiency of data collection. Exposures of up to 300 seconds (limited mostly due to light pollution) are easily achieved at all telescope orientations with stellar profiles exhibiting negligible trailing artefacts.

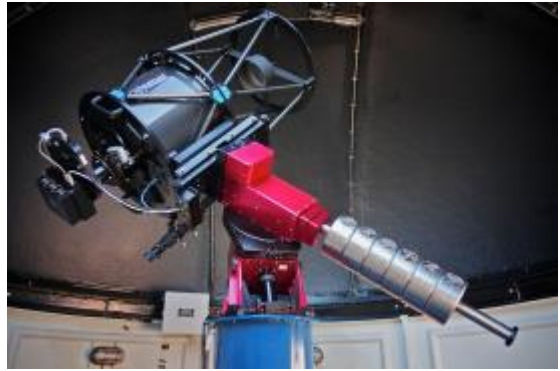

**Figure 2: The SPIRIT II telescope.**

### 2.1.1. Imaging System

SPIRIT I utilises a 0.35m f/11 Schmidt-Cassegrain telescope paired with an Apogee Alta U6 CCD camera incorporating a front illuminated Kodak KAF-1001e sensor with an array of 1024x1024 pixels of 24μm size. This combination provides a square field of view of 20 arc minutes and a native resolution of 1.2 arc seconds per pixel, considered ideal for the sky conditions at UWA. SPIRIT II includes a more sophisticated 0.43m f/6.8 Corrected Dall-Kirkham telescope manufactured by Planewave Instruments. A Finger Lakes Instruments Proline camera utilising a Kodak KAF-16801 sensor with an array of 4096 x 4096 pixels of 9μm size provides a square field of view of 40 arc minutes and a resolution of 0.63 arc seconds per pixel at bin 1. As a consequence, SPIRIT II is nominally used at bin 2 or bin 3 to provide for more appropriate stellar sampling under the skies at UWA. Both systems have been optimised to produce stellar profiles that meet the 2-3 pixel FWHM ideal. The CCD cameras employ non anti-blooming CCD sensors, and as a result exhibit excellent linearity up to saturation.

The SPIRIT I and II image trains include software controlled filter wheels that provide access to photographic red, green and blue filters, and photometric B, V and R filters sourced from Astrodon. A clear (non-IR blocked) filter is provided on both instruments to provide for parfocal 'non-filtered' imaging. SPIRIT I also includes a narrow band H-a filter with a 5nm bandwidth at 656.3nm.

The image train on SPIRIT I is supported by an Optec TCF-S3 temperature controlled focuser which provides a reliable means of maintaining focus throughout the night via a temperature probe attached to the side of the aluminium telescope tube. Initial focus measurements undertaken over a range of temperatures determined a highly linear relationship between temperature and focus position, with the resulting coefficient permanently programmed into the focuser's control software. Only occasional, seasonal re-focusing is necessary. The carbon fibre truss construction of the SPIRIT II telescope provides a thermally stable support for the optics and rarely requires focus adjustment over the course of a typical observing year.

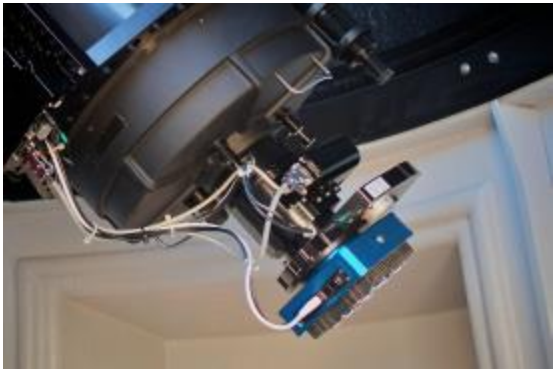

**Figure 3: SPIRIT I optical train, showing the Optec TCF-S3 temperature controlled focuser and Apogee U6 CCD camera and filter wheel.**

It has been difficult to precisely define the magnitude limits of both instruments owing to the constantly changing local sky conditions and pollution gradients over the city of Perth. Using stacking techniques, measurements of a magnitude 19.5 minor planet were achieved with the 0.35m SPIRIT I telescope and successfully submitted to the Minor Planet Centre in 2011. Photometric studies by students of brighter than magnitude 16 stars with exposures of 120s or less routinely yield magnitude errors of around 0.03 or better, as determined by software for targets whose brightness is well above sky background and read noise levels.

## 2.2 Software

The SPIRIT software technical design is centred around an implementation of ASCOM compliancy and utilises ACP Observatory Control Software to provide interoperability between control applications. ACP also includes components necessary for the provision of web-enabled access, fundamental to the SPIRIT design, and at time of writing is still the only commercial observatory control suite with built in web functionality. Users can access and control the SPIRIT telescopes using any modern browser on both desktop and mobile platforms.

The SPIRIT ACP web interface has been customised significantly in a number of areas including:

- Disabling buttons that allow users to undertake unnecessary or undesirable actions, such as opening or closing the dome, modifying the CCD camera's cooling state and viewing non-essential sidebar tools and superfluous information.
- Streamlining the look, feel and text to cater for an audience age range from school students to adults.
- Adding functionality, such as a webcam view, a link to the bookings calendar, enhancing the default help information, and in the case of SPIRIT II, access to the 'SPIRIT Bright Star Spectroscope' (see section 7.1).

A secure, single server model is employed in each of the SPIRIT observatories, with all control software, web server and user file systems installed on a single PC. Back-end interoperability is achieved through commercially available telescope and camera control software as well as a number of ASCOM components which require only minimal adaptation in order to engineer robust unattended dusk until dawn operations. A commercially available telescope sequencing application (CCD Commander) has been employed to automate daily start-up and shutdown tasks invisible to the end user. Functions such as sun-angle timed camera cooling, dome opening, dusk calibration frame acquisition and activation of the ACP web interface are encapsulated within an easily modified daily script. These daily scripts can also be daisy-chained to allow a repeating multi-day operation for weekend use.

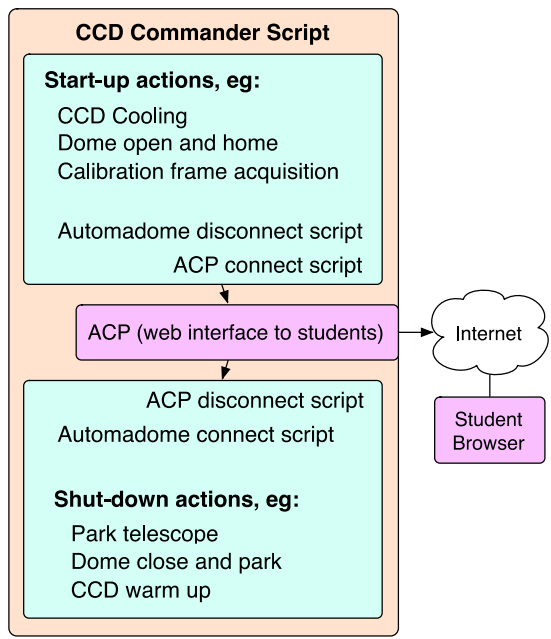

**Figure 4:** Daily script flow. The entire pre-dusk until post-dawn daily sequence is embedded within a CCD Commander 'action file'.

Remote administration is provided using a third-party cloud based solution (GoToMyPC) which enables non-device and non-location dependant administration. SPIRIT can be administered via smart phone from anywhere in the world!

Users acquiring images with SPIRIT can download both RAW and calibrated FITS files directly from within the web interface or using File Transfer Protocol (FTP), which also provides access to log files and calibration frames for advanced analysis. A high quality JPEG version of each image is also automatically generated at the completion of each acquisition. This provides students with a level of instant satisfaction as images appear on the screen before them, as well as an efficient means to monitor image quality without the need to download large FITS formatted image files. Moreover, the JPEG version has proven sufficient for the vast majority of basic student engagement activities including colour astrophotography.

### 2.3 Modes of Operation

Central to the SPIRIT philosophy is a novel approach to cater for a varied student and user demographic—encompassing a range of abilities from primary school students to post graduate tertiary students and researchers. The web interface provides an intuitive choice of operating modes providing a pathway for students to develop more skills and embark on more research-oriented projects.

'Mode 1' allows novice students to 'drive' the telescope, acquiring either a single image, or a series of images through different filters of a single target. This is undertaken in real time, while observing telescope and dome movements together with image acquisition progress via the web interface. 'Mode 2' provides the more experienced student with multi-target unattended options. It is most suited to survey work, and is typically utilised by experienced users of SPIRIT undertaking long term monitoring of variable stars, minor planet astrometry or other survey work. Mode 2 also provides an alternate means of servicing collaborative or large class work—where many students submit observation requests which are coordinated by a teacher or group leader.

A third mode of operation uses *Tools for Automated Observing* (TAO)—a fully automated user acquisition system, including submission and scheduling and operates as a customised script within the ACP control software environment. It ensures maximum telescope utilisation through the use of an advanced scheduler at the expense of real time interaction. It is, essentially, an image request system. Perhaps as a consequence of this, it has not been used within the educational outreach and engagement model characterised by the SPIRIT philosophy.

For the vast majority of students, mode 1 continues to offer an engaging and reliable means of acquiring images. For improved data gathering efficiency and more advanced student science projects such as the *SPIRIT light curve photometry project* (see section 4), mode 2 provides for fully unattended operations. In both cases, students can monitor operations regardless of whether in direct control or running an automated plan.

### 3. Telescope Access

Access to SPIRIT is provided through the SPIRIT web site (spirit.icrar.org) which also presents a 'base of operations' for all things

SPIRIT. The site contains a description of the program, access to guides, documents and activities, image galleries and a news blog. The telescope access page provides authorised users with a clearly defined 5 step booking process with links to weather information, calendar and booking form. Access to the web interface for each telescope is also provided on the access page.

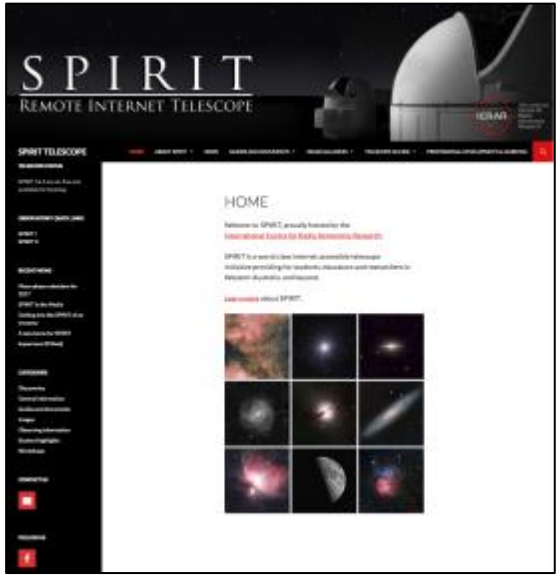

**Figure 5:** The SPIRIT initiative home page.

### 3.1 Telescope Web Interface

The web interface for each telescope provides a modified ACP web presentation including both content and control. Imaging options, together with file access, system status information, a web cam view and other items are contained within a logical side bar menu structure, allowing the users to customise the view according to their requirements.

Targeting objects is facilitated by an object database contained within the ACP software. It contains a subset of the most common deep sky objects so that novice users need only input the catalogue name and select "get coordinates" to target these objects. Similarly, ACP has been configured to facilitate targeting of moving objects such as planets, asteroids and comets—the latter two through an updated minor planet orbital file downloaded periodically and providing updated ephemerides of recently observed or discovered objects. Moving object ephemerides are automatically calculated at the time of imaging for such objects specified in the targeting dialogue.

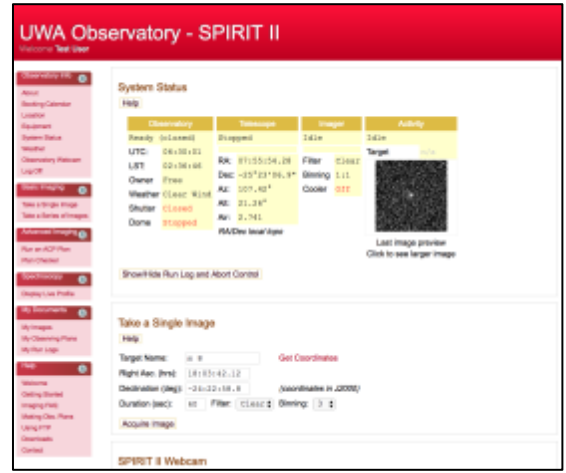

**Figure 6:** The SPIRIT II telescope web interface.

The flexibility of the advanced imaging options allows for fully unattended, multi-target imaging using a range of exposure and filter options. Utilising ACP's intuitive directive-based format, sophisticated text based observing plans can be uploaded and run at the time of booking. The advanced imaging mode allows experienced students to embark on more sophisticated observing programs, such as high cadence variable star photometry observations or 'all night' target monitoring.

### 4. Education Program

The education program has evolved from an initial teacher-centred program to a student-centred program. In-school or on site at UWA workshops are offered for small groups of students embarking on school-based programs. These usually start with an engagement cycle of activities, where students are introduced to the mechanics of the night sky, learn how to remotely access and use the SPIRIT telescopes, acquire images and undertake basic image processing. Packaged as "SPIRIT 101", these activities provide an engaging way for students to learn about contemporary observational astronomy before embarking on more challenging activities. The SPIRIT 101 program of activities has also shown great promise in increasing mainstream participation by less motivated and low achieving students.

With the completion of SPIRIT 101, a number of advanced activities become available:

- *Targeting and imaging asteroids with SPIRIT.* Students learn how to target and observe visible minor planets using the SPIRIT telescopes.

- *Astrophotography with SPIRIT.*

A workshop covering both basic and advanced image processing techniques, enabling users to create stunning colour images of astronomical objects using the SPIRIT telescopes.

- *Advanced astrometry with SPIRIT.*

An advanced minor planet workshop intended for upper school investigations and advanced studies. This workshop provides the tools and techniques necessary to undertake minor planet astrometry and submit observations to the Minor Planet Centre for publication.

- *Variable star photometry with SPIRIT.*

This comprehensive program of activities provides detailed information on how to undertake photometric observations of short period variable stars using SPIRIT. Workshop sessions introduce students to the same tools and techniques used by professional astronomers to create light curve photometry that is submitted to the American Association of Variable Star Observers (AAVSO) database.

All SPIRIT activities are supported by on-line guides and manuals and offered at no cost as part of the ICRAR outreach mandate. Where practical, a visit to the SPIRIT observatory at UWA and ICRAR research institute including presentations by leading astronomers as a component of the outreach program is also encouraged.

Advanced activities often form a component of existing academic extension programs in what is an astronomy-poor state and federal science curriculum in Australia. Remote and regional schools are uniquely positioned to take advantage of SPIRIT's web-enabled interface and tailored learning programs are available to service regional schools at a small cost.

## 4.1 State Curriculum

The results of pilot teacher professional development activities undertaken early in the program informed the development of curriculum specific activities for teachers planning to utilise SPIRIT as part of the astronomy components of the then Western Australian state curriculum. These resources were made available as part of a secondary science teacher enrichment program jointly funded by the WA Department of Education and UWA. These resources are separately available for download through the department's portal.

## 5. Results

### 5.1 Technical Implementation

Reproducibility of the technical design itself was tested early in the program during the commissioning of SPIRIT II—a near exact 'copy and paste' of the SPIRIT I deployment. Aside from variations in device dependencies and their configuration, software installation was substantially achieved by restoring a backup of the SPIRIT I observatory server directly onto the new SPIRIT II server.

Reliability of the original technical design has exceeded expectations, with both the SPIRIT I and SPIRIT II telescopes operating essentially 24x7x365 and experiencing no significant or costly hardware failures over the past 6 years. The original Windows PCs are still in use in both observatories despite considerable temperature and humidity fluctuations over a typical observing year. The Sirius and MaxDome hardware in particular have proven surprisingly robust. Notable issues and replacements include:

- Ice formation on the sensor of the sealed Apogee Alta CCD housing requiring servicing by vendor.
- Replacements fans for both CCD cameras (a near annual requirement).
- Replacement or repair of both Boltwood cloud sensors (continuous working life ~ 2 years).
- Replacement of the computer motherboard in SPIRIT I observatory PC.
- Minor corrosion on exposed USB cables
- Replacement of the dome shutter's 12V DC sealed lead acid batteries every 2 years (precautionary).

In general, it has been found that instrumentation and computer devices are less likely to be problematic if left powered on—with most items at their most vulnerable during cold start up. The solid-state circuitry employed in modern computers, telescope mounts and cameras and in the dome automation used by SPIRIT does not appear to suffer by being left in a powered-on state. Even the modest sized solar panel used to maintain dome shutter battery charge has proven sufficient with the dome shutter mechanism left powered on 24x7.

The use of commercially available components has helped mitigate risks associated with proprietary in-house developed solutions with replacement items available 'off-the-shelf' and often delivered within a short time frame. A modest ~ $5,000 AUD annual maintenance budget for the two facilities has proven adequate over the current operating life. The use of commercially available components has also helped to 'future-proof' the initiative with respect to staff changes, given the ubiquitous use of many of the components in the SPIRIT technical solution, in both professional and advanced amateur installations.

### 5.2 Education and Science

SPIRIT has proven a welcome tool in providing for student engagement in science as is routinely affirmed through the positive feedback received from teachers and an increasing annual user base. Both basic and advanced (colour) imaging has proven a popular means of capturing interest and introducing basic astronomical concepts, such as object types and classification, distance scales and the basics of astronomical data acquisition.

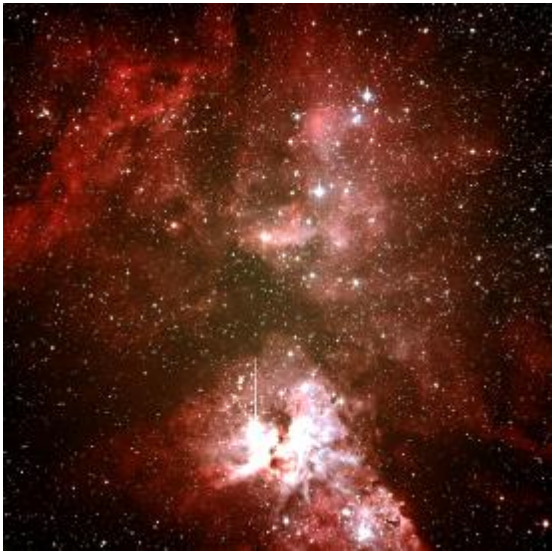

**Figure 7: Eta Carina Nebula, SPIRIT II. Credit: Alex King, John XXIII College Perth, 2014.**

The potential for student participation in 'real science' was demonstrated conclusively in the attainment of UWA's Minor Planet Center observatory code by local high school students during the commissioning of SPIRIT I. Further afield, Students at a school in Japan successfully determined the rotation period of minor planet (15552) sandashounkan in 2015, after an extensive observation program using SPIRIT. Their results, together with others, appear in the IAU's *Minor Planet Center* publications.

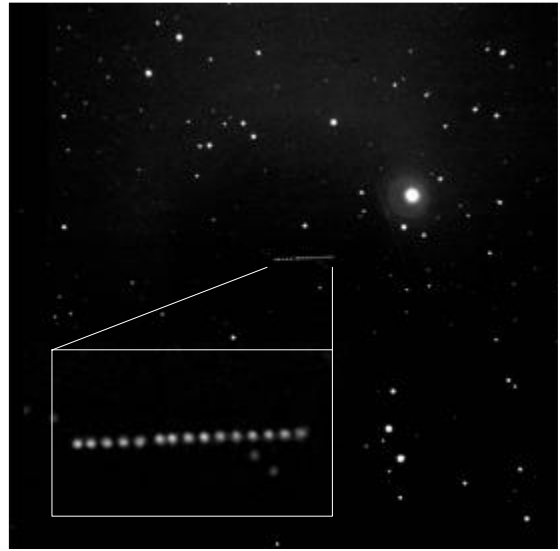

**Figure 8: Time lapse composite image of minor planet 909 Ulla, SPIRIT I. Credit: Gurashish Singh, Mt Lawley Senior High School Perth, 2010.**

At part of an annual 'Girls in STEM' initiative, students from Iona Presentation College in Perth use SPIRIT to undertake observations of RR Lyrae stars, analysing and preparing light curves of these short period variables for submission to the AAVSO. Many of the southern stars observed by these students have little in the way of published photometry and as such provide yet more opportunities for students to undertake novel observational work.

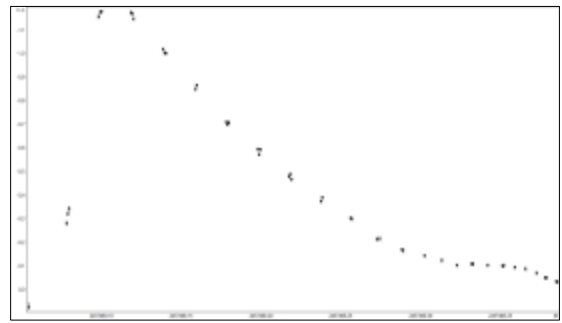

**Figure 9: High school student light curve of FX Hya, a type RR Lyrae variable star in the constellation of Hydra. Credit: Victoria Wong, Iona Presentation College Perth, 2016.**

SPIRIT also enjoys successful use in undergraduate and postgraduate study and research. As part of coursework, UWA astrophysics students use SPIRIT to acquire data of stellar clusters in order to create colour magnitude diagrams using crowded field photometry techniques. Students as far afield as the UK enrolled in on-line post graduate courses in astrophysics have also used SPIRIT to acquire data for Masters level projects.

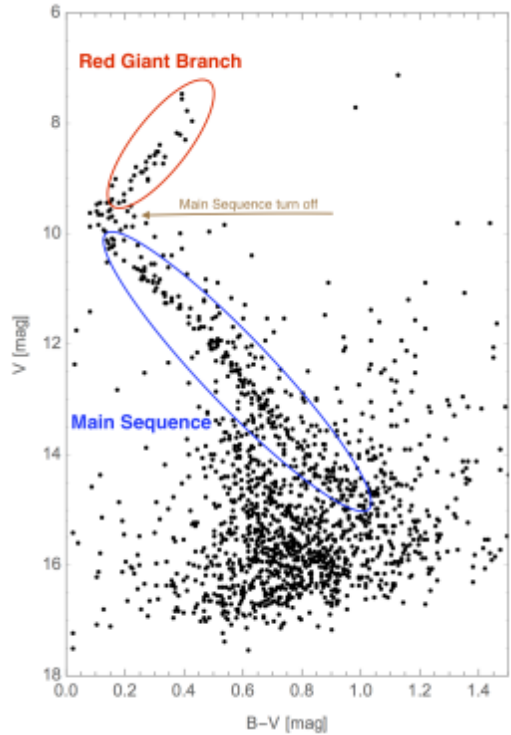

Figure 10: Crowded field photometry using SPIRIT: a colour magnitude diagram of the cluster M93. The main sequence and turn-off points are clearly visible. Credit: Adam Watts, UWA, 2015.

While still in its infancy, the publication of spectroscopic observations from prototype testing of both high and low resolution spectrographs has affirmed the enormous potential that spectroscopy offers for advanced student projects in the coming years (see section 7.1 for more information).

## 6. Lessons Learned

Not withstanding the availability of commercial robotic telescope hardware and software, successfully implementing a diverse suite of hardware and software components benefits from prior expertise in the field. Marrying precise electro-mechanical control with contemporary software platforms and the requirements of a diverse user demographic is a non-trivial endeavour. The term 'robotic' should not be confused with 'easy' and a percipient understanding of the potential for problems should not be underestimated, particularly for continuous operation scenarios. The scalability from single owner/operator to a multi-user environment requires careful consideration, particularly in areas of client interface, access control and user management, as these concepts generally lay outside the domain of commercial telescope control software offerings.

Consideration should also be given to the skill set requirements, useful in the construction, implementation and maintenance of an advanced Internet telescope. These include:

- the ability to operate hand tools and possessing general electrical and mechanical adeptness particularly during construction phases;
- proficient project management skills and a well-developed capacity to solve complex technical problems in areas often characterised by the terms 'cutting edge' or 'bleeding edge';
- experience in astronomy including a working knowledge of the observational techniques used in contemporary CCD equipped telescope implementations;
- understanding of astronomical software design as well as proficiency and experience in the implementation of robotic telescope technologies.

Fostering good relationships in an industry represented by a comparatively small number of commercial vendors is considered advantageous. Likewise is the ability to participate effectively in the broader community of advanced amateurs and professionals who represent the robotic telescope fraternity. In any cutting edge field, achieving success in troubleshooting difficult problems is more often a result of 'who you know' rather than 'what you know'.

Not withstanding these atypical aspects, it is nevertheless the view of this author that this type of project represents both a viable and realistic undertaking given enough impetus.

Of note has been the realisation of the time required to maintain and manage both the telescope facility and the educational program itself. While funding for capital investment in telescope hardware may be comparatively easy to justify, securing on-going funding for staff and program initiatives is much harder.

In a similar vein has been the realisation that without a mature and well produced education program, use of the SPIRIT telescopes within the current educational environment in Western Australia would be significantly reduced. Expertly maintaining a well-structured educational program has proven the most fundamental requirement of the SPIRIT initiative.

## 7. Future Enhancements

### 7.1 Spectroscopy

The provision for web-enabled spectroscopy has long been a consideration for enhanced use and a number of projects have been commissioned in order to identify the requirements and characterise the issues. A small web enabled, low resolution spectroscope was added to SPIRIT II in late 2014. The 'SPIRIT Bright Star Spectroscope' (SBSS) consists of an 80mm refracting telescope, a 100 lines/mm transmission grating and a video camera. It produces a real time, low resolution spectral image of bright stars with a spectral dispersion of approximately 10 Angstroms per pixel. The entire spectral image—from ultraviolet to infrared—is contained in the field of view. Video from the spectroscope is processed in real time and presented as a spectral profile in a customised view created in the ACP web interface. Together with a snapshot of the live spectral profile, the original spectral image can be downloaded for calibration and analysis off-line.

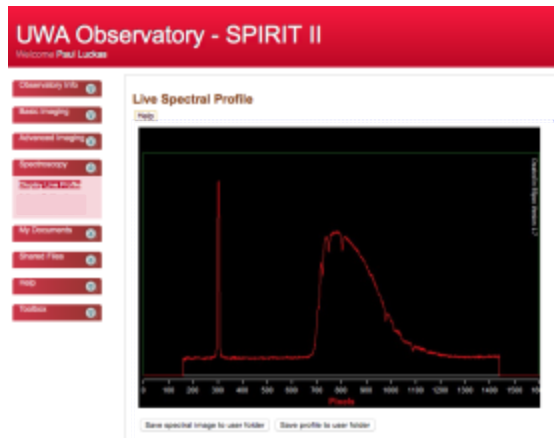

**Figure 11: The uncalibrated spectral profile of a type-A star as it appears in real time on the SPIRIT II web interface. The zero order stellar image appears at far left.**

Students using the SBSS extend their photometric knowledge of the universe into the realm of spectroscopy, learning about the temperature profiles and chemical signatures of stars, and reproducing the foundational work of those such as Edward Pickering and his team at Harvard in the early 20th century.

Deployment of the SBSS was facilitated by software created in-situ, and depends heavily on the functionality of the imaging software used, in this case *RSpec*. As a consequence of this and limitations of control and observatory software, the solution is still prototypical in nature and requires further development.

The scientific capabilities of a high resolution (R > 10000) slit spectrograph on small telescopes under urban conditions have also been tested in a number of configurations since 2013. These tests continue to yield surprising results, affirming the potential for spectroscopy and its tolerance for challenging sky conditions. Pro-am collaborations have resulted in over 500 spectra submitted for publication on a number of massive and 'exotic' star campaigns including those associated with BRITE satellite monitoring and the 2014 eta Carinae periastron campaign (Teodora et al, 2016).

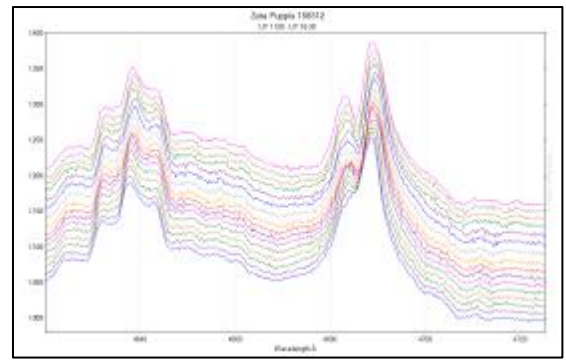

**Figure 12: High resolution time series of the blue supergiant zeta Puppis. Changes in the HeII emission feature can be detected in just a few hours of spectroscopic observation. 257 spectra spanning several months have been included as part of a major study into the large-scale wind structures, due for publication in 2017. Similar high cadence spectra of other massive stars including, gam2 Vel, mu Sgr, uw Cma and WR 6 have produced publication quality spectra as part of other professional collaborations.**

In addition to high-resolution spectroscopy, two galactic novae in 2016 were confirmed in testing of a low resolution (R ~ 600) slit spectroscope in late 2016 (Luckas, 2016).

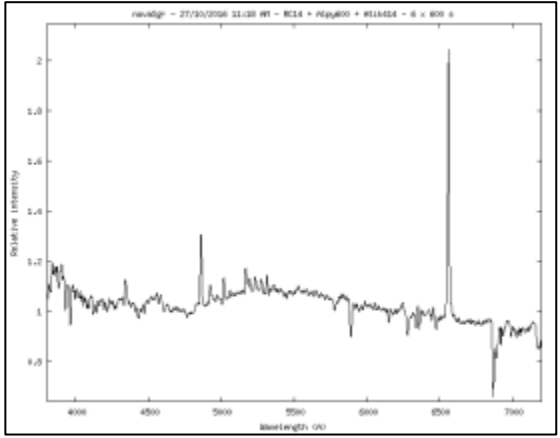

**Figure 13: Confirmation spectra of galactic nova ASASSN-16ma acquired using a Shelyak Alpy600 spectrograph (Luckas, 2016).**

These publications—a number still in draft—allude to the enormous potential for spectroscopy as a means for students to contribute at the very cutting edge of contemporary astronomy using modest sized robotic telescopes equipped with spectrographs and operating under challenging urban skies.

### 7.1.1. Technical Hurdles

Transforming in situ use of a spectrograph into a web enabled solution requires significant modification of the current commercial slit spectroscope offerings. The basic operational requirements for slit spectroscope control include:

- The ability to centre a target on the slit in real time.
- Telescope mount auto-guiding capability.
- Remote control of the calibration arc lamp.
- Adjustment of the grating angle to the target wavelength of interest.
- Collimator focus.

A number of prototype solutions for robotising the above functions have been developed, including *Arduino* control of the Ne arc lamp in two slit spectrographs tested to date. Video monitoring through the spectrograph's conventional auto-guiding port has proven successful as a means of positioning bright targets and the slit, though this requires a separate telescope auto-guiding capability. Rudimentary designs for servo controlled micrometre and collimation lens controls have also been completed. However, commercially available observatory control software does not provide well for multi-device and multi-instrument package scenarios. Limitations presented by the MaxIM DL imaging application within the ACP control environment in particular, have prevented progress in this area. A more plausible solution using a fibre-fed echelle spectrograph, such as the *Shelyak eShel* within a separate image control environment is likely to present the best option for medium term web-enablement.

Figure 14 shows a prototype multi-instrument package including both high and low resolution spectrographs and a photometric CCD camera for conventional imaging. All instruments are selectable via a 'flip mirror'. The auto-guiding ports on both spectrographs have been repurposed with video feeds to aid in remote positioning of the target on the slit. Auto-guiding is accomplished separately using a pick off mirror integrated into the flip mirror device. Care has to be taken to ensure back focus requirements of the telescope design are adhered to, and that the image planes for all sensors are, ideally, at focus.

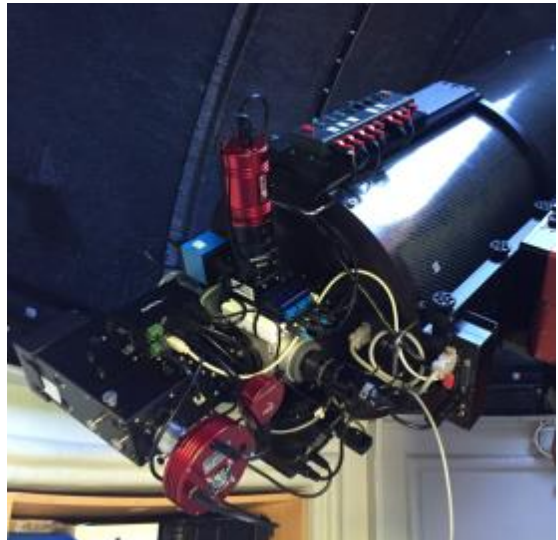

**Figure 14: The author's home observatory and SPIRIT test bench. The multi-instrument optical train includes both high and low resolution spectroscopes, and a conventional CCD imager for photometry.**

This solution has been rigorously tested in number of spectrographic and photometric applications, and can be remotely controlled only using direct remote control software. Web enabling this technology for mainstream multi-audience use remains a goal for the future.

## 7.2 SPIRIT Radio

Multi-wavelength astronomy, and in particular those that include radio wavelengths, is of significant interest to the outreach goals at ICRAR. Radio astronomy presents an exciting opportunity for student observation as it extends into day light (class time) hours. Preliminary testing of a small 'Haystack' design radio telescope at UWA's *Gingin Observatory* located approximately 1 hour north of Perth in a [relative to UWA's Crawley Campus] radio-quiet zone, has provided baseline information on the capabilities of a 1.8m small radio telescope. Projects include measuring the temperature of the sun's photosphere, and detecting HI regions in the galactic plane and in other galaxies. The system is currently only controlled using remote access software, but plans for an in-house developed web interface are being explored.

## 8. Conclusion

The SPIRIT initiative has helped to show that the use of well-designed and maintained contemporary web-enabled, robotic telescopes in astronomy education can be both engaging and relevant. That these telescopes need not be multi-million dollar facilities to provide opportunities for students to participate in authentic and cutting edge science appears also to be well established. The development and delivery of educational support is considered fundamental to the effectiveness of such initiatives, and when done well can do much to create high levels of student participation and engagement in science.

## 10. Additional Figures and Tables

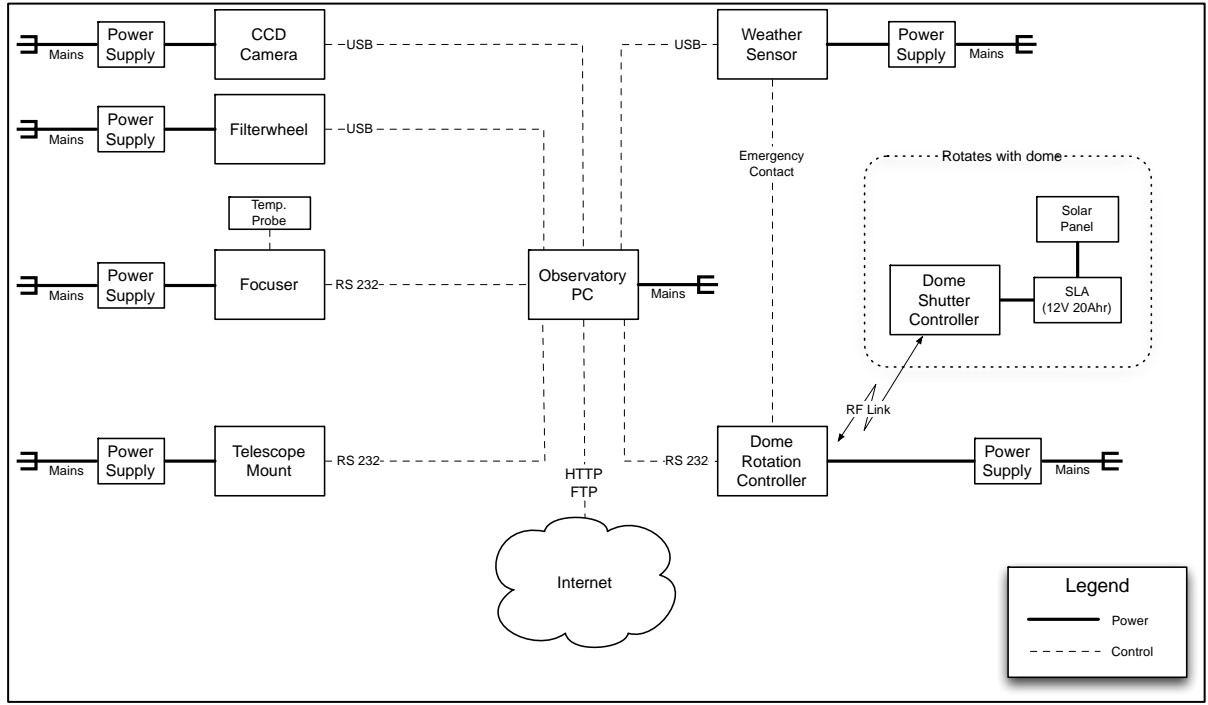

**Figure 15: SPIRIT Hardware Block Diagram**

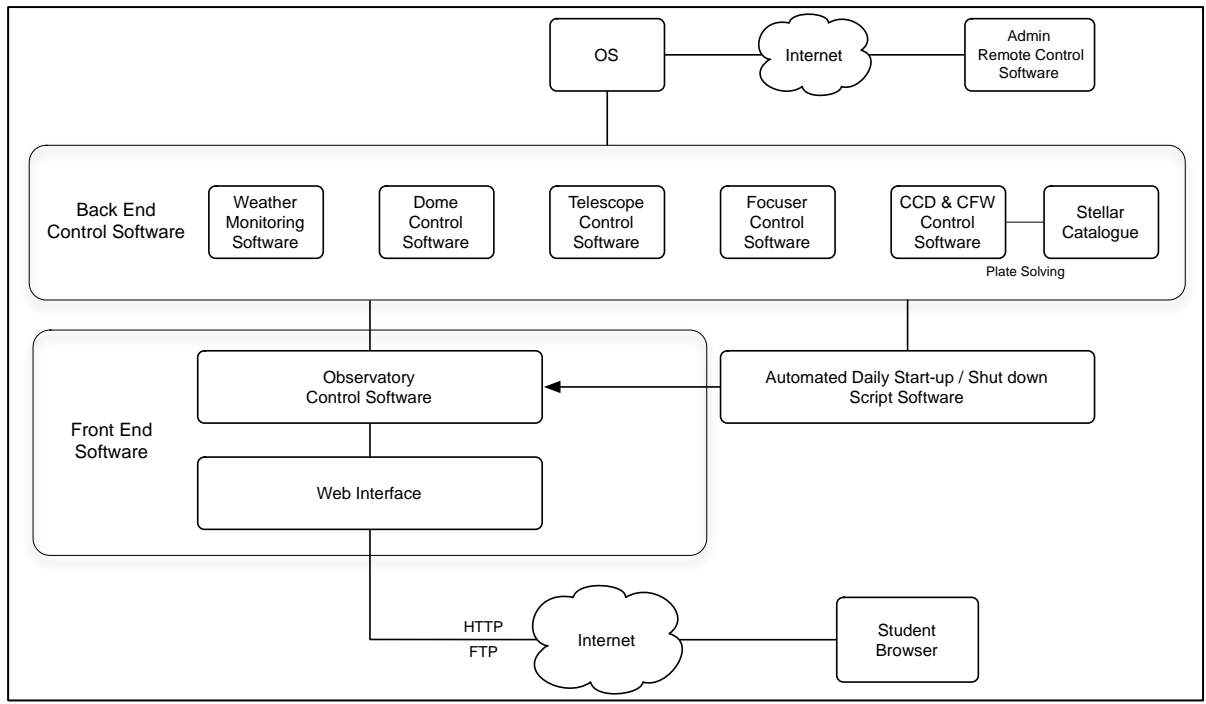

**Figure 16: SPIRIT Software Block Diagram**

| Device | Model | Vendor |
| --- | --- | --- |
| Telescope Mount | Paramount ME | Software Bisque |
| Telescope OTA (SPIRIT I) | C14 Schmidt-Cassegrain | Celestron |
| Telescope OTA (SPIRIT II) | CDK 17 | Planewave Instruments |
| CCD Camera (SPIRIT I) | Alta U6 | Andor (formerly Apogee) |
| Filter wheel (SPIRIT I) | AFW50-9R | Andor (formerly Apogee) |
| CCD Camera (SPIRIT II) | Proline PL-16801 | Finger Lakes Instrumentation |
| Filter wheel (SPIRIT II) | CFW5-7 | Finger Lakes Instrumentation |
| Filters | CRGB + BVR + H-a | Astrodon |
| Focuser SPIRIT I | TCFS-3 | Optec Inc. |
| Dome enclosure | 2.3m & 3.5m models | Sirius Observatories |
| Dome automation | MaxDome | Diffraction Limited |
| Weather monitoring | Boltwood Cloud sensor II | Diffraction Limited |

**Table 1: SPIRIT major hardware items and vendor list.**

| Component | Software | Vendor |
| --- | --- | --- |
| Weather monitoring | Clarity II | Diffraction Limited |
| Dome control | MaxDome | Diffraction Limited |
| Telescope mount control | TheSkyX | Software Bisque |
| Focuser control (SPIRIT I) | FocusMax | CCDWare |
| Focuser control (SPIRIT II) | Planewave Interface | Planewave Instruments |
| CCD and filter wheel control | MaxIM DL | Diffraction Limited |
| Observatory control software | ACP Observatory Control | DC-3 Dreams |
| Web interface | ACP Observatory Control | DC-3 Dreams |
| FTP Server | BulletProof FTP | BulletProof Software |
| Backup software | Acronis True Image | Acronis International GmbH |
| Remote administration | GoToMy PC | Citrix Systems, Inc. |

**Table 2: SPIRIT primary software list.**